\documentclass{eas}
\usepackage{graphicx}
\begin{document}

\title{A census of ultra-compact dwarf galaxies in nearby galaxy clusters} 
\runningtitle{A census of UCDs in nearby galaxy clusters}
\author{Michael Hilker}\address{European Southern Observatory, 
Karl-Schwarzschild-Str.\,2, 85748 Garching b. M\"unchen, Germany,
\email{mhilker@eso.org}}
\begin{abstract}
Ultra-compact dwarf galaxies (UCDs) are predominatly found in the cores of
nearby galaxy clusters. Besides the Fornax and Virgo cluster, UCDs have
also been confirmed in the twice as distant Hydra\,I and Centaurus clusters.
Having (nearly) complete samples of UCDs in some of these clusters allows the 
study of the bulk properties with respect to the environment they are living 
in. Moreover, the relation of UCDs to other stellar systems in galaxy 
clusters, like globular clusters and dwarf ellipticals, can be investigated 
in detail with the present data sets. The general finding is that UCDs seem 
to be a heterogenous class of objects. Their spatial distribution within the 
clusters is in between those of globular clusters and dwarf ellipticals. In 
the colour-magnitude diagram, blue/metal-poor UCDs coincide with the sequence 
of nuclear star clusters, whereas red/metal-rich UCDs reach to higher masses 
and might have originated from the amalgamation of massive star cluster 
complexes in merger or starburst galaxies.
\end{abstract}
\maketitle
\section{Introduction}
Galaxy clusters are known to be the environments that harbour the largest
number of early-type dwarf galaxies. Despite dwarf ellipticals (dEs) and 
dwarf spheroidals (dSphs), a new type of old, compact stellar systems with 
stellar masses in the range of faint dwarf galaxies ($\sim10^7 M_{\odot}$)
was identified in nearby galaxy clusters, the so-called ultra-compact dwarf 
galaxies (UCDs) (Hilker et al. \cite{hil99}, Drinkwater et al. \cite{dri00}).
Although very similar to globular clusters (GCs) in many respects, UCDs show 
two main differences to them: they follow a mass-size relation (Ha\c{s}egan 
et al. \cite{has05}) and show an elevated mass-to-light ratio in comparison 
to GCs of the same metallicity (Mieske et al. \cite{mie08a}). The latter 
finding cannot be explained by simple stellar population models with a 
canonical IMF (Dabringhausen et al. \cite{dab08}, Mieske \& Kroupa 
\cite{mie08b}).  Both characteristics are valid for UCDs with masses 
exceeding $\sim 2\times10^6 M_{\odot}$ which corresponds to a luminosity 
limit of about $M_V<-11$ mag.
Often this luminosity is used to divide UCDs from globular clusters. However,
this does not mean that there might exist fainter UCDs and brighter GCs. 
This depends on the still unknown formation mechanisms of both types. One 
might think of $\omega$ Centauri and M\,54 as faint UCDs.

The main formation scenarios for UCDs can be divided into a galaxian and a
star cluster origin. The former assume that UCDs are either the remnant
nuclei of disrupted galaxies (e.g. Bekki et al. \cite{bek01}) or genuine
compact dwarf galaxies (e.g. Phillipps et al. \cite{phi01}). The latter
include the formation of UCDs in stellar supercluster complexes of merging 
galaxies (Fellhauer et al. \cite{fel02}) or via `normal' GC formation in
giant molecular clouds (Mieske et al. \cite{mie02}).
Photometric and structural properties alone are not sufficient to decide
whether UCDs are of star cluster or galaxian origin (see for a review
Hilker \cite{hil09}).

A promising way to learn more about the nature of UCDs is to study their
global properties in galaxy clusters and compare them to those of other dwarf
galaxies and rich globular cluster systems around central cluster galaxies.

\section{Samples of UCDs in nearby galaxy clusters}

Finding UCDs is not a straightforward task. They can easily be confused with
foreground stars and unresolved background galaxies. It's like searching
needles in a haystack. Indeed, the first UCDs were discovered not until about
ten years ago by systematic spectroscopic surveys in nearby galaxy clusters, 
aiming at confirming the membership of either galaxies or globular clusters
(Hilker et al. \cite{hil99}, Drinkwater et al. \cite{dri00}). Only now we can
say that statistically meaningful samples are becoming available.

In Table\,1 the current number counts of UCDs and GCs/UCDs more luminous
than $\omega$\,Cen in various clusters are listed. The Fornax cluster, so far,
is the best 
studied environment with $\sim60$ UCDs and $>150$ $\omega$\,Cens. Within its
core radius the completeness is of the order of 90\%. For the other clusters
this value is much lower. In the Virgo cluster data on UCDs still are quite 
sparse (e.g. Jones et al. \cite{jon06}), but probably will outnumber all 
other UCD samples when the Next Generation Virgo Cluster Survey and its 
spectroscopic follow-up will become available.

\begin{table}
\caption{Number of UCDs (Cols. 3-5) and GCs/UCDs brighter than $\omega$\,Cen
(Cols. 6 and 7) within different cluster-centric radii as indicated in the
second row. The last column gives the estimated completeness of UCD counts 
within one core radius.}
\vskip0.2cm
\begin{tabular}{lccccccc}
\hline
Cluster & $R_{\rm core}$ & \multicolumn{3}{c}{$N_{\rm UCD}$} & 
 \multicolumn{2}{c}{$N_{>\omega \rm Cen}$} & $C_{\rm UCD}$ \\
 & [kpc] & all & $<100$kpc & $<0.5R_c$ & all & $<R_c$ & $<R_c$\\
\hline
Fornax    & 100 & 59 & 34 & 20 & 154 & 106 & $\sim90\%$ \\
Hydra I   & 350 & 38 & 31 & 26 &  65 &  56 & $<70\%$ \\ 
Centaurus & 220 & 28 & 20 & 22 & ... & ... & $<40\%$ \\
Virgo     & 350 & $>25$ & $>25$ & $>20$ & ... & ... & ... \\
\hline
\end{tabular}
\end{table}

\section{Spatial distribution of UCDs}

\begin{figure}
\begin{center}
\includegraphics[width=11.5cm]{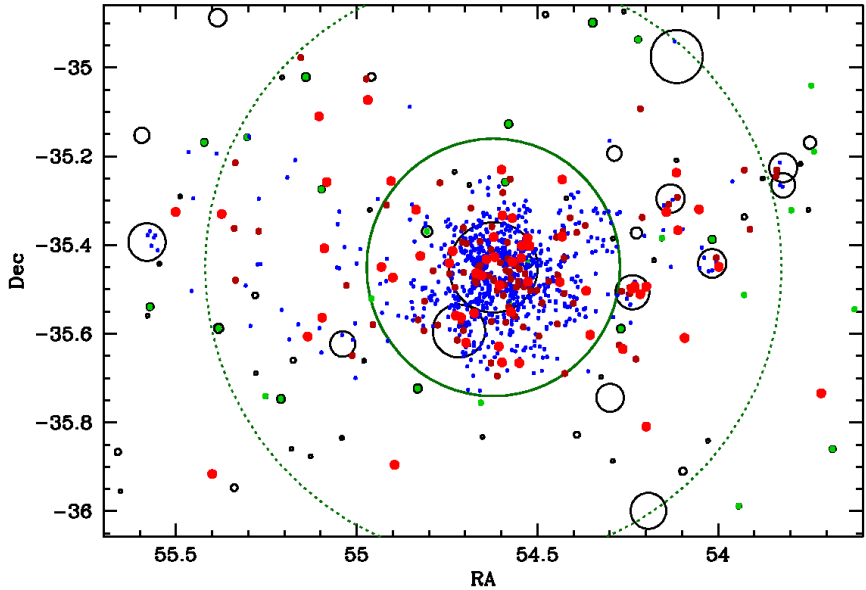}
\caption{The central region of the Fornax cluster. The dotted circle 
indicates the core radius. All objects in this plot are confirmed cluster
members. The black circles are the cluster galaxies, nucleated dEs are 
marked with a green dot. The small blue dots are GCs with $M_V<-8.5$ mag,
small red dots $\omega$\,Cen-like GCs ($M_V<-10.4$ mag), and large red 
dots UCDs ($M_V<-11$ mag). Objects within the dark green circle were used to
derive the cumulative radial number distributions of the different objects
(see Fig.\,2).}
\end{center}
\end{figure}

Based on the UCD samples available, one can study their spatial distribution
within areas of high number count completeness. As an example, in Fig.\,2
the location of GCs, UCDs, nucleated dEs and other galaxies in the central
region of the Fornax cluster are shown. All of them are confirmed members,
eother by radial 
velocity measurements or morpholocical classification (in the case of 
galaxies). As can be seen, many of the UCDs are concentrated around the 
central cluster galaxy NGC\,1399, but also around other major cluster 
galaxies. Those seem to belong to the globular cluster populations of the
individual host galaxies. On the other hand, there exists a fair number of 
UCDs in the intra-cluster space, well beyond the tidal radii of the major 
cluster galaxies. Those cannot easily be explained by a `normal' globular 
cluster population, but rather belong to the cluster population.

When plotting the cumulative radial number distribution of the different
stellar systems and galaxies within 110 kpc (the 90\% complete area), a
clear sequence of concentration becomes apparent (see Fig.\,2). From high to 
low concentration the order is: red (metal-rich) GCs, blue (metal-poor) GCs,
GCs/UCDs more luminous than $\omega$\,Cen ($M_V<-10.4$ mag), UCDs ($M_V<-11$
mag), nucleated dEs, and finally all cluster galaxies. Although UCDs clearly
do not share the same distribution as dE,Ns, they also are not as 
concentrated as globular clusters. This again demonstrates that UCDs seem
to be a mixed bag of objects, some being related to the central globular
cluster system, others most probably being related to a galaxian formation
process, i.e. they might be stripped nuclei of dEs.

\begin{figure}
\begin{center}
\includegraphics[width=8.0cm]{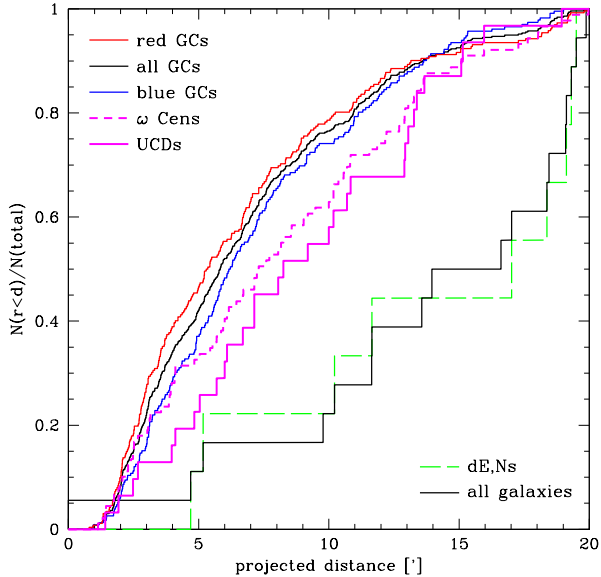}
\caption{Cumulative radial number distributions of globular clusters, UCDs,
dE,Ns and galaxies within 110 kpc of the central cluster galaxy NGC\,1399.}
\end{center}
\end{figure}

The cumulative radial number distributions of UCDs in the other galaxy
clusters presented in Table 1 show the same behaviour -- despite smaller 
number counts: UCDs are more concentrated towards the cluster centre than the
cluster galaxies, but less concentrated than the central globular cluster
systems.

\section{UCDs in the context of other hot stellar systems}

\begin{figure}
\begin{center}
\includegraphics[width=9.5cm]{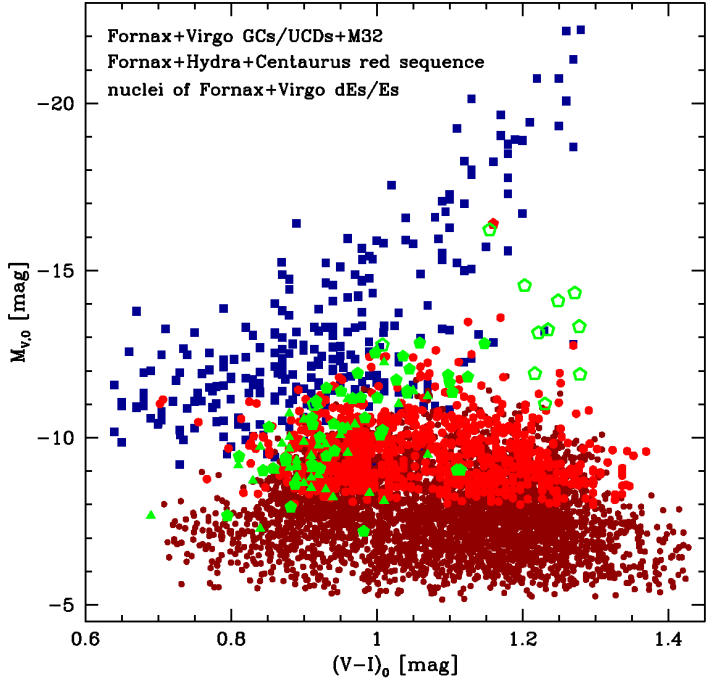}
\caption{Colour-magnitude diagram of various stellar systems in the nearby 
galaxy clusters Fornax, Virgo, Hydra\,I and Centaurus. Blue squares mark the
red sequence of early-type galaxies (Mieske et al. \cite{mie07}, Misgeld et 
al. \cite{mis08}, \cite{mis09}). The small, dark red dots are GCs around the 
central galaxies in Virgo and Fornax from the ACS surveys (Peng et al. 
\cite{pen06}, Jord\'an et al. \cite{jor09}). Light red dots are confirmed 
GCs/UCDs in Fornax (Schuberth et al. \cite{sch10}). The red filled hexagon 
marks M32. The green hexagons and triangles are nuclear star clusters (NCs) 
of early-type galaxies in Fornax and Virgo (Lotz et al. \cite{lot04}, 
C\^ot\'e et al. \cite{cot06}), open symbols those of giant galaxies.}
\end{center}
\end{figure}

The photometric and structural parameters of UCDs can be compared to those
of other hot stellar systems, in particular globular cluster, nuclear star
cluster and early-type dwarf galaxies. In Fig.\,3 the colour-magnitude
diagram of those systems is presented, assembled from data in the Fornax,
Virgo, Hydra\,I and Centaurus clusters. The colour $(V-I)_0=1.05$ mag
divides the blue (metal-poor) GCs from the red (metal-rich) ones, showing the
well-known colour bimodality. The most luminous, and thus most massive UCDs 
extend the red GC population to higher luminosities. Some of those might have 
formed in the same event as the bulk of the red GCs, which are believed to be 
related to the bulge/spheroid formation of ellipticals. A plausible scenario 
for that are major mergers of gas-rich galaxies that are known to create 
stellar supercluster complexes.

UCDs with blue colours, instead, coincide in the CMD with the location of 
nuclear star clusters of early-type galaxies. The colour-magnitude relation
of NCs falls on top of the blue side of the metal-poor GC sequence, also
known as blue tilt (e.g. Harris \cite{har09}). Thus, if UCDs are remnant
nuclei of disrupted dwarf galaxies they might extend to fainter luminosities
and form part of the metal-poor globular cluster systems. The similar 
size-luminosity relation of UCDs and nuclear star clusters is another 
intriguing parallel between both types of objects (see contribution by
Misgeld et al. in this volume).

The ultimative answer about the nature of UCDs can only be given, if the
global, photometric and structural parameters of UCDs are accompanied by
detailed kinematic and spectroscopic abundance analyses.

\section{Conclusions}

Ultra-compact dwarf galaxies are defined through their mass-size relation
and enhanced dynamical mass-to-light ratios -- roughly occuring for stellar
systems with masses $>2\times10^6 M_{\odot}$. UCDs share some properties of
globular clusters as well as of galaxy nuclei, thus they might be an
inhomogeneous class of objects. They are mostly concentrated around major
galaxies but also are found in the intra-cluster space. As a consequence
UCDs are spatially more concentrated than the cluster galaxies but show a
wider distribution than the rich globular cluster systems around the central
cluster galaxies. Still the studies of the UCD population in nearby galaxy
clusters suffer from incompleteness effects. More spectroscopic surveys are
needed to get a full account of UCDs. Large samples of well-studied UCDs
(i.e. with radial velocities, spectroscopic element abundances and age 
estimates), like in the Fornax cluster, offer the opportunity to unravel 
the nature of UCDs via `chemo-dynamical' tagging.


\end{document}